\documentclass[preprint,amsmath,amssymb,aps,showkeys,showpacs]{revtex4}
\usepackage[colorlinks=true, pdfstartview=FitV, linkcolor=blue, citecolor=red, urlcolor=magenta]{hyperref}
\usepackage{graphicx}
\usepackage{latexsym}
\usepackage{amsmath}
\usepackage{amsfonts}
\usepackage{amssymb}
\usepackage{verbatim}
\usepackage{dcolumn}
\usepackage{amssymb}
\usepackage{bm}
\usepackage{float}
\usepackage{subfigure}
\usepackage[english]{babel}
\newcommand{\be}{\begin{equation}}
\newcommand{\ee}{\end{equation}}
\newcommand{\bea}{\begin{eqnarray}}
\newcommand{\eea}{\end{eqnarray}}

\begin{document}

\newcommand{\NITK}{
\affiliation{Department of Physics, National Institute of Technology Karnataka, Surathkal  575 025, India}
}

\title{Repulsive Interactions in the Microstructure of Regular Hayward Black Hole in Anti-de Sitter Spacetime}

\author{Naveena Kumara A.}
\email{naviphysics@gmail.com}
\NITK
\author{Ahmed Rizwan C.L.}
\email{ahmedrizwancl@gmail.com}
\NITK
\author{Kartheek Hegde}
\email{hegde.kartheek@gmail.com}
\NITK
\author{Ajith K.M.}
\email{ajith@nitk.ac.in}
\NITK

\begin{abstract}
We study the interaction between the microstructures of Hayward-AdS black hole using Ruppeiner geometry. Our investigation shows that the dominant interaction between the black hole molecules is attractive in most part of the parametric space of temperature and volume, as in van der Waals system. However, in contrast to the van der Waals fluid, there exists a weak dominant repulsive interaction for small black hole phase in some parameter range. This result clearly distinguishes the interactions in a magnetically charged black hole from that of van der Waals fluid. However, these sort of interactions are characteristic for charged black holes since they do not dependent on magnetic charge or temperature.
\end{abstract}


\maketitle

\section{Introduction}
In recent years the subject of black hole chemistry has become a window to probe the properties of AdS black holes. In black hole chemistry, to study the phase transitions of AdS black holes, the negative cosmological constant of the AdS spacetime is identified as the thermodynamic variable pressure \cite{Kastor:2009wy, Dolan:2011xt}. Interestingly, the phase transition of certain AdS black holes analytically resemble that of a van der Waals system \cite{Kubiznak2012, Gunasekaran2012, Kubiznak:2016qmn}. Recently, by studying the phase transitions, attempts were made to investigate the underlying microscopic properties of the AdS black holes \cite{Wei2015, Wei2019a, Wei2019b, Guo2019, Miao2017, Zangeneh2017, Wei:2019ctz, Kumara:2019xgt, Kumara:2020mvo, Xu:2019nnp, Chabab2018, Deng2017, Miao2019a, Chen2019, Du2019, Dehyadegari2017, Ghosh:2019pwy, Ghosh:2020kba}. In these studies, the geometric methods were the key tools in probing the microscopic details of the black holes. In contrast to the statistical investigation of ordinary thermodynamics, the approach here is in the reverse order, the macroscopic thermodynamic details are the ingredients for the microscopic study \cite{Ruppeinerb2008}.  This technique is inspired by the applications of thermodynamic geometry to ordinary thermodynamic systems \cite{Ruppeiner95, Janyszek_1990, Oshima_1999x, Mirza2008, PhysRevE.88.032123}. In Ruppeiner geometry, the sign of the curvature scalar indicates the nature of interaction between the constituent particles, negative for attractive and positive for repulsive interactions. Besides that, the curvature scalar diverges near the phase transition point. Therefore, the correlation length can be associated with the Ruppeiner scalar, which gives more insights into black hole microstructures.

Recently, a general Ruppeiner geometry framework was developed from the Boltzmann entropy formula, to study the black hole microstructure \cite{Wei2019a}. The fluctuation coordinates were taken as the temperature and volume, and a universal metric was constructed. When this methodology was applied to the van der Waals fluid, only a dominant attractive interaction was observed, as it should be. However, when the same methodology is used for the RN-AdS black hole, a different result was obtained. A repulsive interaction between black hole molecules was found in a small parameter range, in addition to the dominant attractive interaction in the rest of the parameter space \cite{Wei2019a, Wei2019b}. Even so, in the case of five-dimensional neutral Gauss-Bonnet black hole, only a dominant attractive interaction was discovered, which is similar to van der Waals fluid \cite{Wei:2019ctz}. Therefore, in general, the nature of the black hole molecular interactions are not universal. In our recent work \cite{Kumara:2020mvo}, we have observed that there exists a repulsive interaction in Hayward-AdS black hole, like that in the RN-AdS case. In the present work, we will make a detailed investigation of the previously observed repulsive interaction.

The primary motivation for our research is due to the great interest regarding regular black holes in black hole physics, as they do not possess any physical singularities. Wide variety of regular black holes exist, ranging from the first solution given by Bardeen \cite{Bardeen1973}, and its modifications \cite{AyonBeato:1998ub, AyonBeato:2000zs}, to the one in which we are interested, the Hayward black hole \cite{Hayward:2005gi}. (We suggest the readers to go through our previous articles  \cite{A.:2019mqv, Kumara:2020mvo} for a detailed discussion on this). Hayward black hole is a solution to Einstein gravity non-linearly coupled to an electromagnetic field which carries a magnetic charge. The secondary motivation for the present study comes from the question, what is the nature of the microstructure of a black hole with magnetically charged constituents? In this article, we probe the phase structure and repulsive interactions in the microstructure of this magnetically charged AdS black hole.

The article is organised as follows. In the following section, we discuss the phase structure of the black hole. Then the Ruppeiner geometry for the black hole is constructed for microstructure study (section \ref{sectwo}). Then we present our findings in section \ref{secthree}.


\section{Phase structure of the Hayward-AdS Black Hole}
\label{secone}
The Hayward black hole solution in a four dimensional AdS background is given by \cite{Fan:2016hvf, Fan:2016rih} (see Ref.  \cite{Kumara:2020mvo} for a brief explanation), 
\begin{equation}
ds^2=-f(r)dt^2+\frac{dr^2}{f(r)}+r^2d\Omega ^2,
\end{equation} 
where $d\Omega ^2=d\theta ^2+\sin \theta ^2d\phi ^2$, and the metric function,
\begin{equation}
f(r)=1-\frac{2 M r^2}{g^3+r^3}+\frac{8}{3} \pi  P r^2.
\end{equation}
We study the phase structure in the extended phase space, where the cosmological constant $\Lambda$ gives the pressure term  $P=-\Lambda /8\pi$. The parameter $g$ is related to the total magnetic charge of the black hole $Q_m$ as,
\begin{equation}
Q_m=\frac{g^2}{\sqrt{2\alpha}},
\end{equation}
where $\alpha$ is a free integration constant. The thermodynamic quantities; temperature, volume and entropy of the black hole are easily obtained as,
\begin{equation}
T=\frac{f'(r_+)}{4\pi}=\frac{2 P r^4}{g^3+r^3}-\frac{g^3}{2 \pi  r \left(g^3+r^3\right)}+\frac{r^2}{4 \pi  \left(g^3+r^3\right)}; \label{temperature}
\end{equation}
\begin{equation}
V=\frac{4}{3} \pi  \left(g^3+r^3\right) \quad \textit{and} \quad S=2 \pi  \left(\frac{r^2}{2}-\frac{g^3}{r}\right).
\end{equation}
These results are consistent with the first law,
\begin{equation}
dM=TdS+\Psi dQ_m+VdP+\Pi d \alpha,
\end{equation}
and the Smarr relation,
\begin{equation}
M=2(TS-VP+\Pi \alpha)+\Psi Q_m.
\end{equation}
The heat capacity of the black hole system at constant volume is,
\begin{equation}
C_V=T\left( \frac{\partial S}{\partial T}\right)_V=0.
\label{cv}
\end{equation}
Inverting the expression for the Hawking temperature (\ref{temperature}) we get the equation of state,
\begin{equation}
P=\frac{g^3}{4 \pi  r^5}+\frac{g^3 T}{2 r^4}-\frac{1}{8 \pi  r^2}+\frac{T}{2 r}. \label{stateeqn}
\end{equation}
From the state equation, we can observe that the black hole shows a critical behaviour similar to that found in a van der Waals system. This is often interpreted as the transition between a small black hole (SBH) and a large black hole (LBH) phases. In our earlier study \cite{Kumara:2020mvo}, we have shown that an alternate interpretation is possible using Landau theory of continuous phase transition, where the phase transition is between the black hole phases at different potentials. In this alternate view, the black hole phases, namely high potential, intermediate potential and low potential phases, are determined by the magnetic charge. In either of these interpretations, the phase transition can be studied by choosing a pair of conjugate variables like $(P-V)$ or $(T-S)$.   With the conjugate pair $(P,V)$, the Maxwell's equal area law has the form,
\begin{equation}
P_0(V_2-V_1)=\int _{V_1}^{V_2}PdV. \label{equalarea}
\end{equation}
Since there exists no analytical expression for the coexistence curve of Hayward-AdS black hole, we usually seek numerical solutions. For that, we obtain the following key ingredient from the Maxwell's equal area law. Using the Eq. (\ref{equalarea}) and  expressions for $P_0(V_1)$ and $P_0(V_2)$ from equation of state (\ref{stateeqn}) we get the solution as $P_0(x)$ and $T_0(x)$  (see Ref. \citep{Kumara:2020mvo}). We have taken $x=r_1/r_2$, where $r_1$ and $r_2$ are the radii of black hole for first-order phase transition points. The critical values are readily obtained by setting  $x=1$,
\begin{equation}
T_{c}=\frac{\left(5 \sqrt{2}-4 \sqrt{3}\right) n_1^{2/3}}{4\times 2^{5/6} \pi  g},
\end{equation}
\begin{equation}
P_{c}=\frac{3 n_2}{16 \times 2^{2/3} n_1^{5/3} \pi  g^2},
\end{equation}
and 
\begin{equation}
V_c=4 n_3 \pi  g^3,
\end{equation}
where $n_1=3\sqrt{6}+7$, $n_2=\sqrt{6}+3$ and $n_3=2\sqrt{6}+5$.
The reduced thermodynamic variables are defined as,
\begin{equation}
T_r=\frac{T}{T_c},\quad P_r=\frac{P}{P_c}, \quad V_r=\frac{V}{V_c}.
\end{equation}
Using these we can write the equation of state in the reduced parameter space,

\begin{equation}
P_r=\frac{2^{2/3} n_1^{5/3} \left[V_r \left(\left(2n_1\right)^{2/3} T_r \left(3 n_3 V_r-1\right)^{1/3}-2n_3\right)+2\right]}{n_2 \left[3 n_3 V_r-1\right]^{5/3}}.
\label{reducedeos}
\end{equation}
The reduced equation of state is independent of the magnetic charge parameter $g$. From the reduced state equation we obtain the spinodal curve, which separates metastable phases from the unstable phase, using the condition,
\begin{equation}
\left( \partial _{V_r} P_r \right)_{T_r}=0.
\end{equation}
The explicit form of the spinodal curve is,
\begin{equation}
T_{rsp}=\frac{ 2^{4/3} n_3 \left[n_3 V_r-2\right]}{n_1^{2/3} \left[n_3 V_r+1\right] \left[3 n_3 V_r-1\right]^{1/3}}.
\end{equation}

\begin{figure*}[t]
\centering
\subfigure[ref1][]{\includegraphics[width=0.425\textwidth]{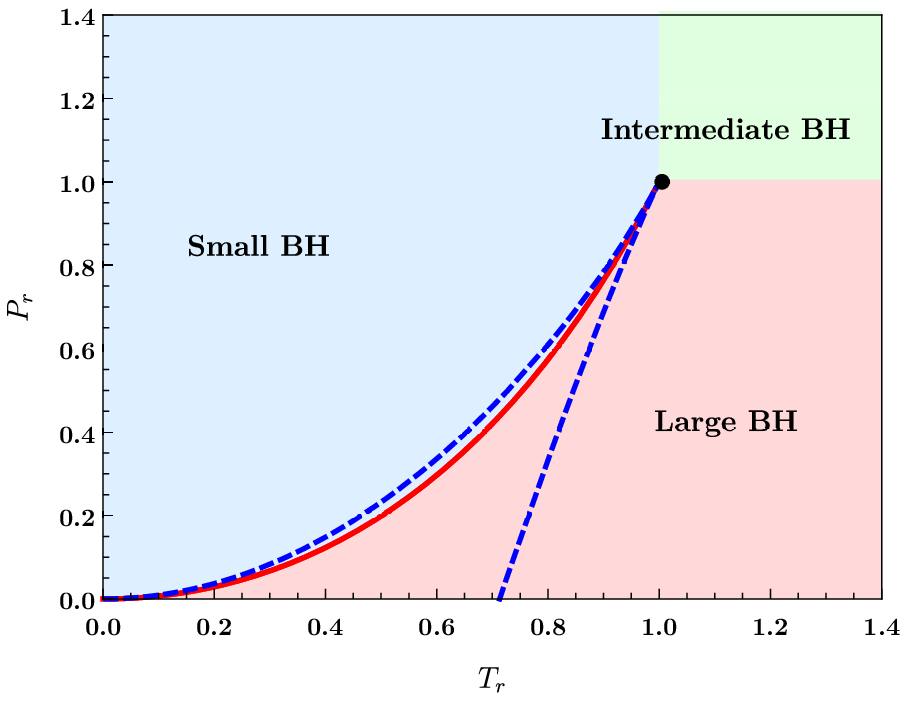}
\label{HPT}}
\qquad
\subfigure[ref2][]{\includegraphics[width=0.425\textwidth]{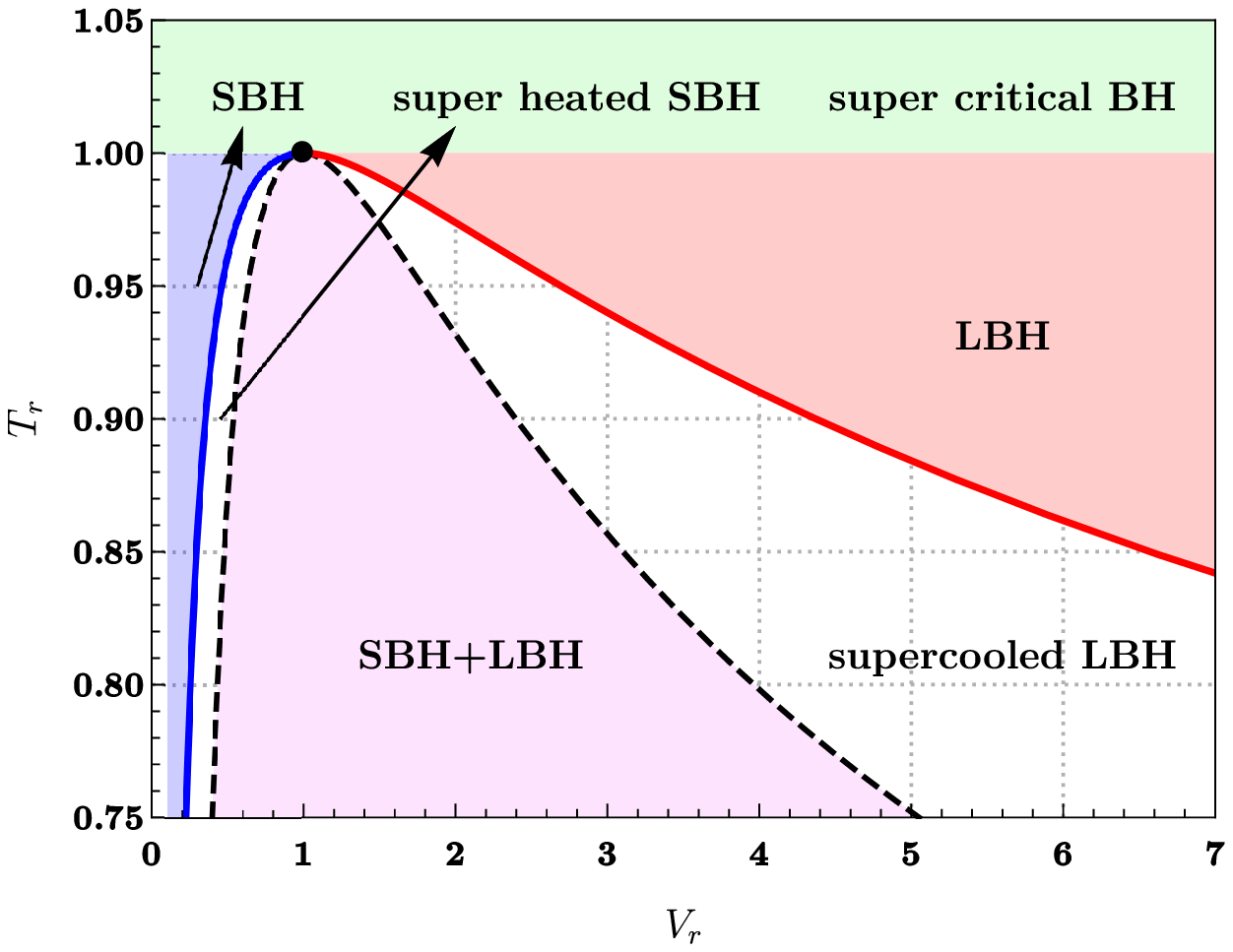}
\label{HTV}}
\caption{Phase structure of Hayward-AdS black hole in $P-T$ and $T-V$ diagrams. The coexistence curve is represented by a solid line and the spinodal curve is shown with a dashed line.  }
\label{fig1}
\end{figure*}

\begin{figure*}[t]
\centering
\subfigure[ref3][]{\includegraphics[width=0.425\textwidth]{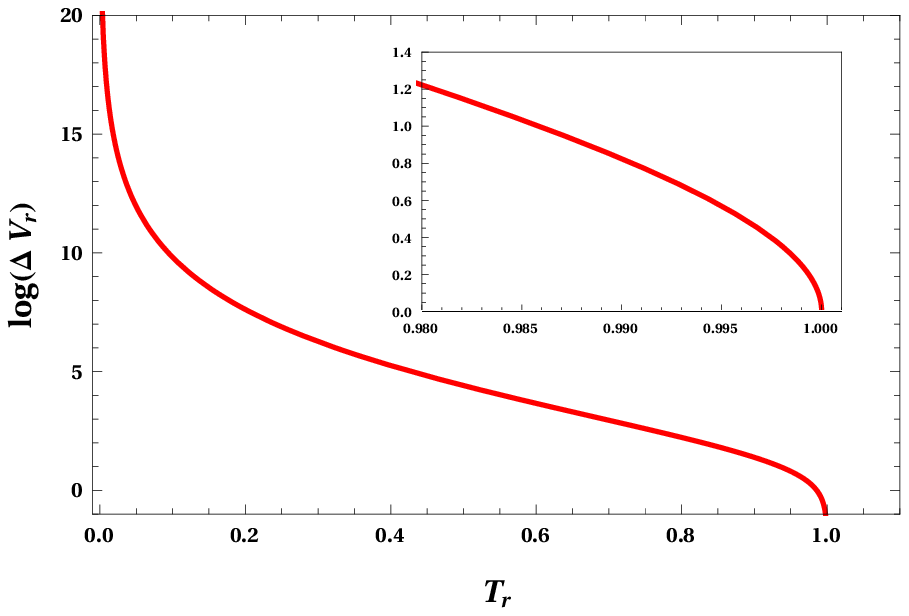}
\label{HVTorder}}
\qquad
\subfigure[ref4][]{\includegraphics[width=0.425\textwidth]{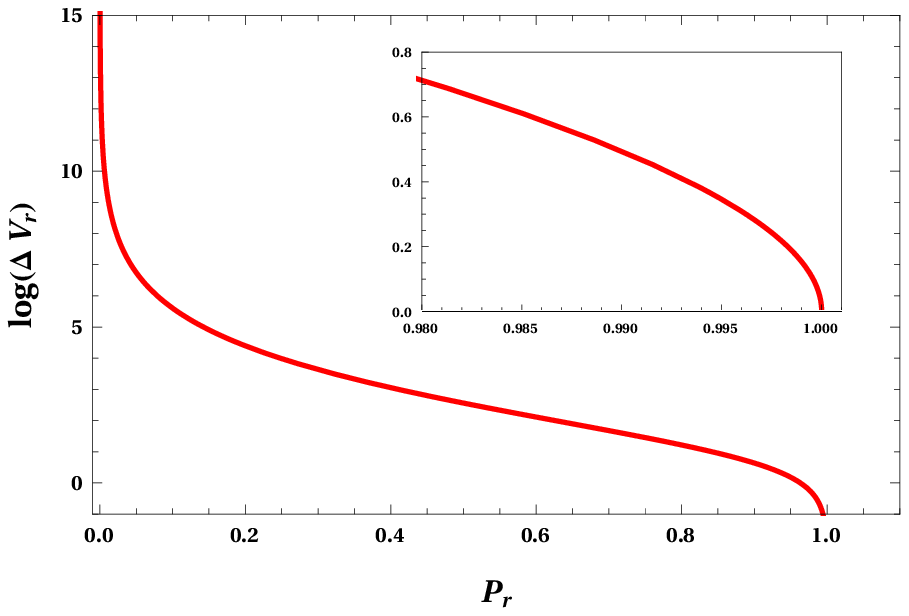}
\label{HVPorder}}
\caption{The behaviour of volume change $\Delta V_r=V_{rl}-V_{rs}$ during phase transition of the black hole. The behaviour near the critical point is shown in inlets using linear scale.}
\label{Deltav}
\end{figure*}

Solving this for $V_r$ and substituting in Eq. (\ref{reducedeos}) we obtain the curve in $P-V$ plane. The spinodal curve along with the coexistence curve display the stable, unstable and metastable phases of the black hole. The coexistence curve is obtained numerically using the expressions $r_2(x)$, $P_0(x)$ and $T_0(x)$. The spinodal and coexistence curves are shown together in Fig. \ref{fig1}. By fitting the coexistence curve in $P-T$ plane we obtained the following expression,

\begin{eqnarray}
P_r=&5.622 \times 10^{-7}-5.539\times 10^{-5} T_r+0.693 T_r^2+0.1365 T_r^3+0.1966 T_r^4\nonumber \\
&-0.4255 T_r^5+1.134 T_r^6-1.698 T_r^7+1.621 T_r^8-0.8651 T_r^9+0.2085 T_r^{10}. 
\end{eqnarray}

The SBH, LBH and supercritical BH phases are depicted in Fig. \ref{HPT}. The coexistence curve separates the SBH and LBH phases. It terminates at the critical point, after which the distinction between the SBH and LBH states is not possible, hence this corresponds to the supercritical black holes. The region between the coexistence curve and spinodal curve corresponds to the metastable states, namely the supercooled LBH and the superheated SBH, which are shown in the $T-V$ diagram (Fig. \ref{HTV}). An observable feature in these diagrams is that the spinodal and coexistence curves meet each other at the critical point.

Now, we would like to study the change in volume during the black hole phase transition as a function of temperature and pressure. Using expression of $r_2(x)$, we make the functional change $V(r)\rightarrow V(x)$ to obtain a parametric expression for $\Delta V_r$. The parametric expression of $\Delta V_r$ along with that of $T_r$ and $P_r$ ($T_0(x)$ and $P_0(x)$ expressions ) are used to plot Fig. \ref{Deltav}, which gives the behaviour of $\Delta V_r$. From Fig. \ref{Deltav} it is clear that, $\Delta V_r$ decreases with increase in both temperature and pressure. It approaches zero at the critical point ($T_r=1$ and $P_r=1$). The numerical calculation reveals that, the behaviour near the critical point is,
\begin{eqnarray}
\Delta V_r = 9.5005 (1-T_r)^{0.5296}, \\
\Delta V_r = 5.4984 (1-P_r)^{0.5216}.
\end{eqnarray}
This suggests that the change in volume $\Delta V_r$ can serve as the order parameter to characterise the black hole phase transition, with a universal critical exponent $1/2$.

\begin{figure*}[t]
\centering
\subfigure[ref5][]{\includegraphics[width=0.425\textwidth]{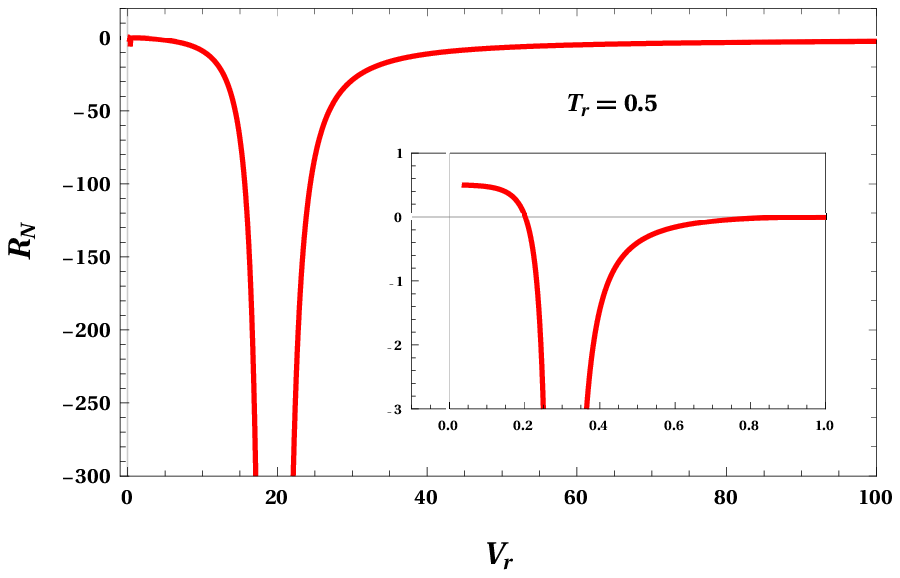}\label{HRNV1}}
\qquad
\subfigure[ref6][]{\includegraphics[width=0.425\textwidth]{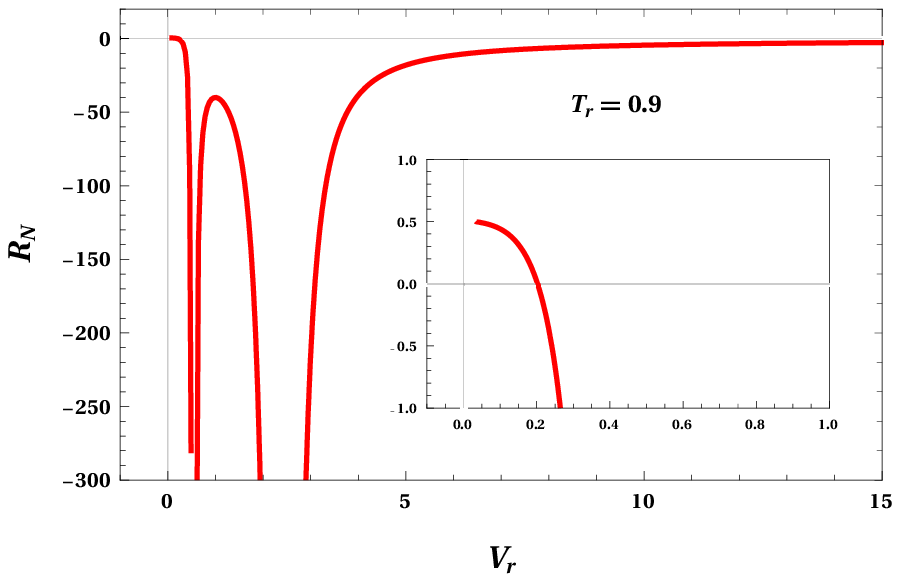}\label{HRNV2}}

\subfigure[ref7][]{\includegraphics[width=0.425\textwidth]{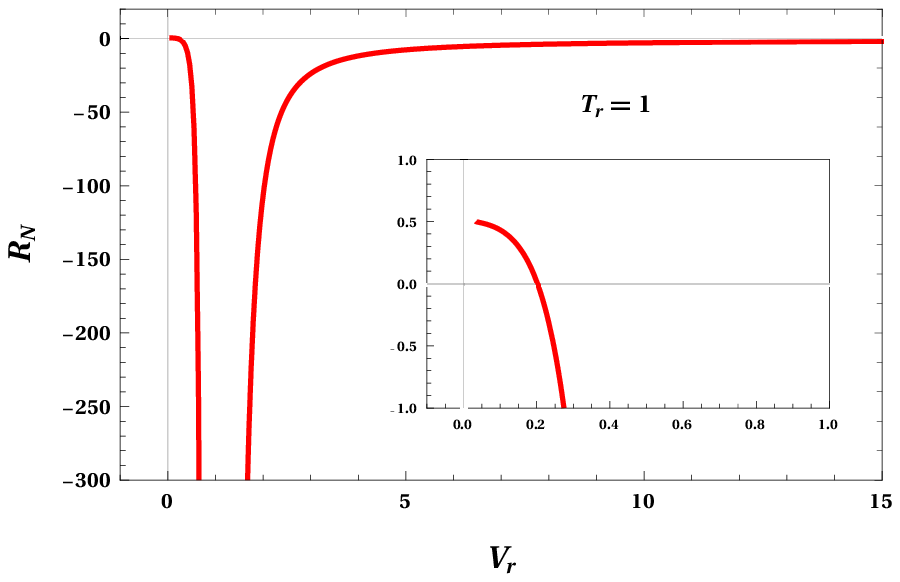}\label{HRNV3}}
\qquad
\subfigure[ref8][]{\includegraphics[width=0.425\textwidth]{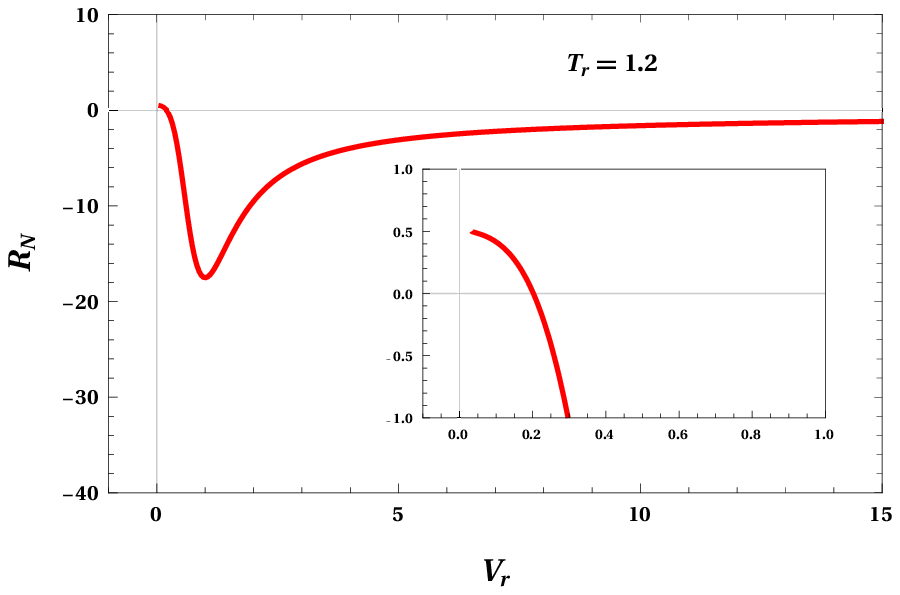}\label{HRNV4}}
\caption{The behaviour of the normalised curvature scalar $R_N$ against the reduced volume $V_r$ at constant temperature. }
\label{RN}
\end{figure*}


\section{Microstructure of the Hayward-AdS Black Hole}
\label{sectwo}

In this section we examine the microstructure of the black hole using Ruppeiner geometry in which $T$ and $V$ are taken as fluctuation coordinates. The line element in this parameter space has the form \cite{Wei2019a},
\begin{equation}
dl^2=\frac{C_V}{T^2}dT^2-\frac{\left(  \partial _V P\right)_T }{T}dV^2.
\label{line}
\end{equation}
The heat capacity $C_V$ vanishes for the Hayward-AdS black hole (Eq. \ref{cv}). This makes the line element (\ref{line}) singular, hence the corresponding geometry will not give the information regarding the microstructure of the black hole. Therefore the normalised scalar curvature is used for studying the microscopic interactions,
\begin{equation}
R_N=C_V R.
\end{equation}
From a straightforward calculation, for the Hayward-AdS black hole we obtain,
\begin{equation}
R_N=\frac{\left[8 \pi  g^3-V\right] \left[ 8 \pi g^3 \left(\tau +1\right)+V \left(2 \tau -1\right)\right]}{2 \left[4 \pi  g^3 \left(\tau +2\right)+V \left(\tau -1\right)\right]^2}.
\end{equation}
with
\begin{equation}
\tau =\pi ^{2/3} T \left(6 V-8 \pi  g^3\right)^{1/3}.
\end{equation}
In terms of the reduced parameters,

\begin{equation}
 R_N= \frac{4 \left[n_3 V_r-2\right]\left[-A+2 n_3 V_r-4\right]}{\left[A-4 n_3 V_r+8\right]^2},
\end{equation}
where,
\begin{equation}
A=2^{1/6} n_1^{2/3} T_r \left(\sqrt{2} V_r+5 \sqrt{2}-4 \sqrt{3}\right) \left(3 n_3 V_r-1\right)^{1/3}.
\end{equation}
Similar to the case of charged AdS black hole and Gauss Bonnet black hole, $R_N$ is independent of $g$. The normalised curvature scalar $R_N$ diverges along the spinodal curve. 

The behaviour of $R_N$ with reduced volume $V_r$ for a fixed temperature is studied in Fig. (\ref{RN}). For $T_r<1$, below critical temperature, $R_N$ has two negative divergence points. They approach each other as the temperature increases and merge together at $V_r=1$ for $T_r=1$. These divergences do not exist for temperatures greater than the critical value. We see that there always exists small regions where the curvature scalar is positive (shown in inlets).  We need to examine whether these regions are thermodynamically stable. Setting $R_N=0$ we get,
\begin{equation}
T_0=\frac{T_{rsp}}{2}=\frac{2^{5/6} \left[n_3 V_r-2\right]}{n_1^{2/3} \left[\sqrt{2} V_r+5 \sqrt{2}-4 \sqrt{3}\right] \left[3 n_3 V_r-1\right]}.
\end{equation}
This is the sign-changing temperature (where curvature scalar changes its sign) , which is half of the spinodal curve temperature as in vdW system, RN-AdS and Gauss-Bonnet black holes. Another solution for this is, $V_r=2/n_3   \equiv V_0.$ 

\begin{figure*}[t]
\centering
\subfigure[ref9][]{\includegraphics[width=0.425\textwidth]{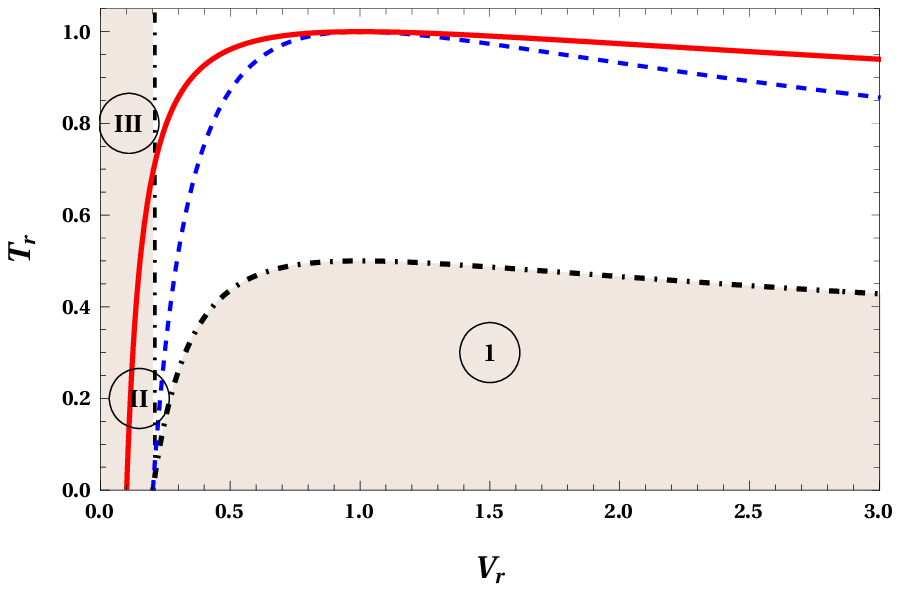}
\label{HSign}}
\qquad
\subfigure[ref10][]{\includegraphics[width=0.425\textwidth]{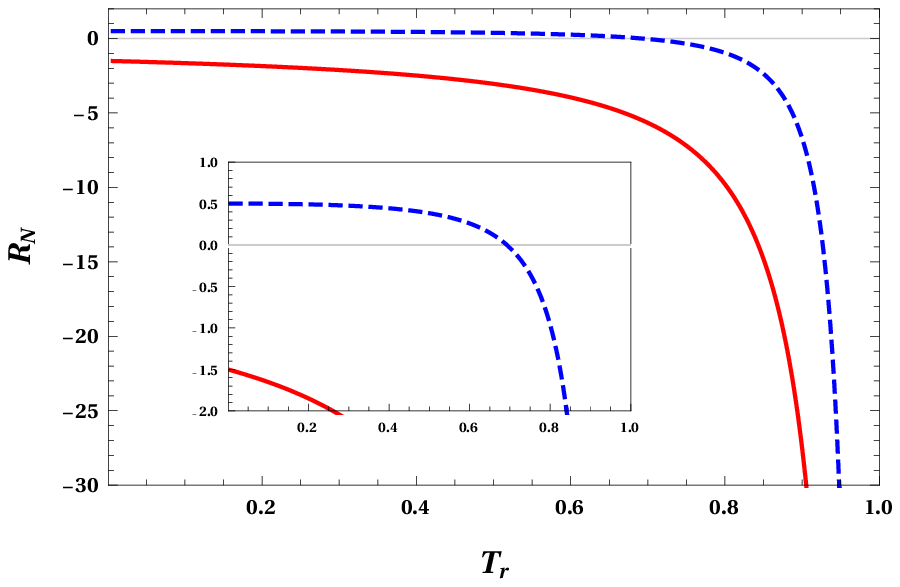}
\label{HRT}}
\caption{ \ref{HSign}: The sign changing curve (dot-dashed black) of $R_N$ along with the coexistence (solid red) and spinodal (dashed blue) curves. The vertical black (dot-dashed) line corresponds to $V_0$. \ref{HRT}: The behaviour of normalised curvature scalar $R_N$ along the coexistence line. The red (solid) line and  blue (dashed) line corresponds to LBH and SBH, respectively. The inlet shows the region where the SBH branch takes positive $R_N$ value.}
\label{FIG4}
\end{figure*}

\begin{figure*}[t]
\centering
\subfigure[ref1][]{\includegraphics[width=0.425\textwidth]{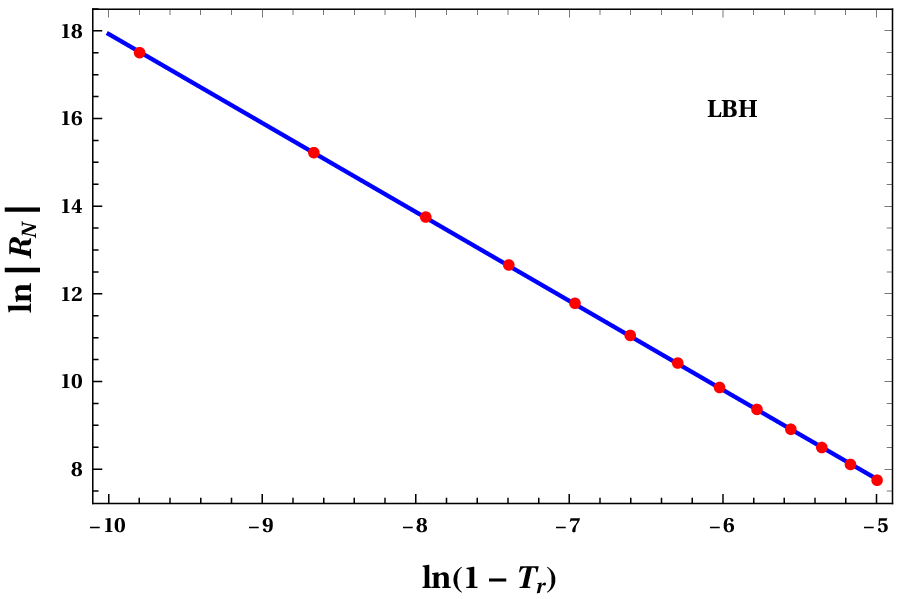}
\label{lbhfit}}
\qquad
\subfigure[ref2][]{\includegraphics[width=0.425\textwidth]{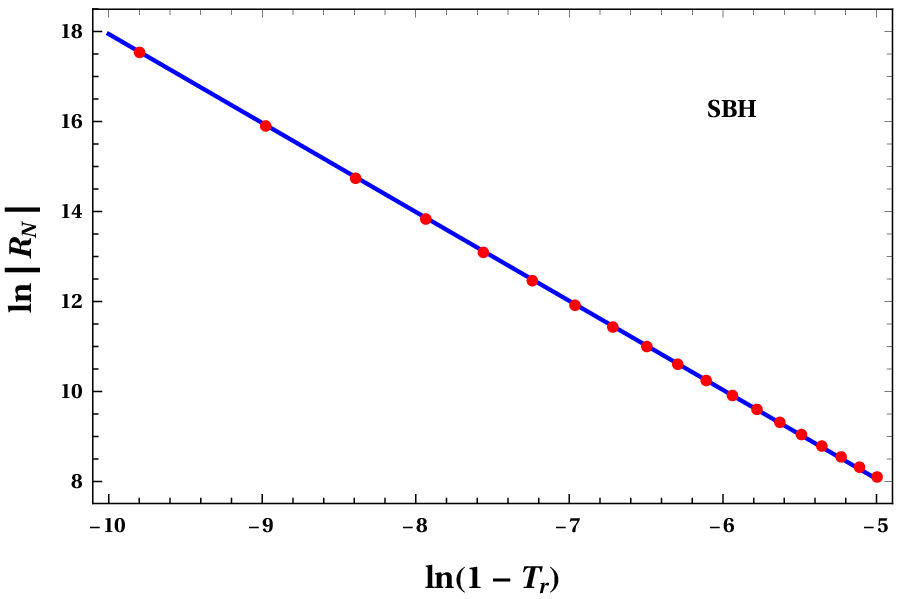}
\label{sbhfit}}
\caption{The numerical fit of $\ln \left|R_N\right|$ vs. $\ln (1-T_r)$ for LBH (left) and  SBH (right) branches. The solid blue line correspond to the plot of fitting formula and the red dots correspond to numerical data.}
\label{fig5}
\end{figure*}

The spinodal, sign changing and coexistence curves are shown together in Fig. \ref{HSign}. The region ($I$) under the spinodal curve for $V_r>V_0$, $R_N$ is positive, which corresponds to the coexistence phase of SBH and LBH, similar to van der Waals fluid's coexistence phase. Everywhere below $V_0$, $R_N$ is positive, including region above and below the coexistence curve. The region ($II$) under the coexistence curve is the same as the previous case, a coexistence phase. However, in the region ($III$) above the curve, which is a SBH phase, we can safely say that the black hole molecules possess repulsive interaction. Therefore in Hayward-AdS black hole, for a small parameter range there exist dominant repulsive interaction. This result is similar to RN-AdS black hole and in contrast to five-dimensional neutral Gauss-Bonnet black hole, where there is no repulsive interaction.

Finally, we consider the behaviour of the scalar curvature $R_N$ along the coexistence curve. Since there exists no analytical expression for the coexistence curve, the analytical study of the curvature scalar behaviour is not possible. The numerical solution is obtained and shown in Fig. \ref{HRT}. Both the SBH branch and LBH branch of $R_N$ have divergences near the critical point. For a large black hole, the sign of $R_N$ is always negative and hence the microscopic interaction is always attractive. Interestingly, for the small black hole, there is a lower temperature range where $R_N$ is positive (inlet of the diagram). This indicates a repulsive interaction between the black hole molecules. From this, we can conclude that in the low-temperature regime, the microstructure, as well as microscopic interaction of the black hole changes drastically during the phase transition. Whereas in the high-temperature range, only microstructure changes, and the nature of interaction remains attractive in both phases. These results are strikingly different from that of van der Waals fluid, where the dominant interaction among the molecules is always attractive.

The critical phenomena associated with the Ruppeiner curvature along the coexistence curve is also studied numerically, as in Ref. \citep{Wei2019b}. The numerical fit is obtained by assuming the following behaviour near the critical point,
\begin{equation}
R_N\sim (1-Tr)^p.
\end{equation}
Which reduces to
\begin{equation}
\ln |R_N|=-p \ln (1-T_r) +q.
\end{equation}
From the numerical fit for the SBH and LBH branches we obtain, 
\begin{eqnarray}
\textit{SBH} :\quad \ln R_N=-1.97733 \ln (1-T_r) -1.83058 \label{sbhnum}\\
\textit{LBH} :\quad \ln R_N=-2.03007 \ln (1-T_r) -2.37298 \label{lbhnum}
\end{eqnarray}
These are depicted in Fig. \ref{fig5} along with the numerical data. The results suggest that $p\approx 2$, which we set as $p=2$ considering the numerical error. Combining Eq. (\ref{sbhnum}) and Eq. (\ref{lbhnum}) we construct,
\begin{equation}
R_N (1-Tr)^2=-e^{-(1.83058+2.37298)/2}=-0.122238 \approx -1/8.
\end{equation}
This agrees with previously obtained results for vdW fluid and other AdS black holes \citep{Wei2019a, Wei2019b, Wei:2019ctz}, that $R_N$ has a universal exponent $2$ and the relation $R_N(1-T_r)^2=-1/8$, near the critical point.


\section{Discussions}
\label{secthree}

In this paper, we have studied the phase transitions and microstructure of the Hayward-AdS black hole. The microscopic properties are analysed from the behaviour of Ruppeiner curvature scalar along the coexistence curve. Since an analytical expression for the coexistence curve is not feasible we have carried out our investigation numerically. In the first part of the paper, we probed the phase structure of the black hole using the coexistence curve in $P_r-T_r$ and $T_r-V_r$ planes. Along with this, spinodal curve is also plotted, which enables us to identify the metastable phases of the black holes, namely the superheated SBH and the supercooled LBH. It is shown that the change in volume $\Delta V_r$ during the SBH - LBH phase transition can serve as an order parameter to describe the same. The behaviour of $\Delta V_r$ has a critical exponent $1/2$ which is universal. 

In the second part of this article, we have focused on the Ruppeiner geometry of the black hole. We have adopted the definition of curvature scalar given in the Ref. \cite{Wei2019a}, where the fluctuation coordinates are temperature and volume. The normalised curvature scalar diverges to the negative infinity at the critical point. Also, we numerically confirm that in the vicinity of critical point $R_N$ has a critical exponent 2 and $R_N(1-T_r)^2\approx -1/8$. Even though the black hole shows van der Waals like phase transition, the microstructure properties differ in some aspects. In van der Waals fluid the dominant interaction among the constituent molecules is always attractive, which does not change during the phase transition. The change in microstructure does not lead to any change in the nature of microscopic interaction. However, in Hayward-AdS black hole there exists a domain, low-temperature range for the small black hole, where the dominant interaction between the black hole molecules is repulsive, which is inferred from the positive sign of the normalised curvature scalar. During the phase transition, in this temperature range, the microscopic interaction of the black hole changes significantly. This result is similar to what is observed in RN-AdS black hole and in contrast to the five-dimensional neutral Gauss-Bonnet black hole, where the interaction is always attractive like van der Waals fluid. To conclude, the magnetic charge in the Hayward-AdS black hole plays a similar role as the electric charge in RN-AdS black hole in contributing to the microstructure. We believe that this is another significant step in understanding black hole microstructure properties. 


\acknowledgments
Authors N.K.A. , A.R.C.L. and K.H. would like to thank U.G.C. Govt. of India for financial assistance under UGC-NET-SRF scheme. Authors would like to thank Shreyas Punacha for his help in numerical calculations.


  \bibliography{BibTex}

\providecommand{\href}[2]{#2}\begingroup\raggedright\begin{thebibliography}{10}

\bibitem{Kastor:2009wy}
D.~Kastor, S.~Ray and J.~Traschen, \emph{{Enthalpy and the Mechanics of AdS
  Black Holes}},
  \href{https://doi.org/10.1088/0264-9381/26/19/195011}{\emph{Class. Quant.
  Grav.} {\bfseries 26} (2009) 195011}
  [\href{https://arxiv.org/abs/0904.2765}{{\ttfamily 0904.2765}}].

\bibitem{Dolan:2011xt}
B.~P. Dolan, \emph{{Pressure and volume in the first law of black hole
  thermodynamics}},
  \href{https://doi.org/10.1088/0264-9381/28/23/235017}{\emph{Class. Quant.
  Grav.} {\bfseries 28} (2011) 235017}
  [\href{https://arxiv.org/abs/1106.6260}{{\ttfamily 1106.6260}}].

\bibitem{Kubiznak2012}
D.~Kubiznak and R.~B. Mann, \emph{{P-V criticality of charged AdS black
  holes}}, \href{https://doi.org/10.1007/JHEP07(2012)033}{\emph{JHEP}
  {\bfseries 07} (2012) 033} [\href{https://arxiv.org/abs/1205.0559}{{\ttfamily
  1205.0559}}].

\bibitem{Gunasekaran2012}
S.~Gunasekaran, R.~B. Mann and D.~Kubiznak, \emph{{Extended phase space
  thermodynamics for charged and rotating black holes and Born-Infeld vacuum
  polarization}}, \href{https://doi.org/10.1007/JHEP11(2012)110}{\emph{JHEP}
  {\bfseries 11} (2012) 110} [\href{https://arxiv.org/abs/1208.6251}{{\ttfamily
  1208.6251}}].

\bibitem{Kubiznak:2016qmn}
D.~Kubiznak, R.~B. Mann and M.~Teo, \emph{{Black hole chemistry: thermodynamics
  with Lambda}}, \href{https://doi.org/10.1088/1361-6382/aa5c69}{\emph{Class.
  Quant. Grav.} {\bfseries 34} (2017) 063001}
  [\href{https://arxiv.org/abs/1608.06147}{{\ttfamily 1608.06147}}].

\bibitem{Wei2015}
S.-W. Wei and Y.-X. Liu, \emph{{Insight into the Microscopic Structure of an
  AdS Black Hole from a Thermodynamical Phase Transition}},
  \href{https://doi.org/10.1103/PhysRevLett.116.169903,
  10.1103/PhysRevLett.115.111302}{\emph{Phys. Rev. Lett.} {\bfseries 115}
  (2015) 111302} [\href{https://arxiv.org/abs/1502.00386}{{\ttfamily
  1502.00386}}].

\bibitem{Wei2019a}
S.-W. Wei, Y.-X. Liu and R.~B. Mann, \emph{{Repulsive Interactions and
  Universal Properties of Charged Anti–de Sitter Black Hole
  Microstructures}},
  \href{https://doi.org/10.1103/PhysRevLett.123.071103}{\emph{Phys. Rev. Lett.}
  {\bfseries 123} (2019) 071103}
  [\href{https://arxiv.org/abs/1906.10840}{{\ttfamily 1906.10840}}].

\bibitem{Wei2019b}
S.-W. Wei, Y.-X. Liu and R.~B. Mann, \emph{{Ruppeiner Geometry, Phase
  Transitions, and the Microstructure of Charged AdS Black Holes}},
  \href{https://doi.org/10.1103/PhysRevD.100.124033}{\emph{Phys. Rev.}
  {\bfseries D100} (2019) 124033}
  [\href{https://arxiv.org/abs/1909.03887}{{\ttfamily 1909.03887}}].

\bibitem{Guo2019}
X.-Y. Guo, H.-F. Li, L.-C. Zhang and R.~Zhao, \emph{{Microstructure and
  continuous phase transition of a Reissner-Nordstrom-AdS black hole}},
  \href{https://doi.org/10.1103/PhysRevD.100.064036}{\emph{Phys. Rev.}
  {\bfseries D100} (2019) 064036}
  [\href{https://arxiv.org/abs/1901.04703}{{\ttfamily 1901.04703}}].

\bibitem{Miao2017}
Y.-G. Miao and Z.-M. Xu, \emph{{On thermal molecular potential among
  micromolecules in charged AdS black holes}},
  \href{https://doi.org/10.1103/PhysRevD.98.044001}{\emph{Phys. Rev.}
  {\bfseries D98} (2018) 044001}
  [\href{https://arxiv.org/abs/1712.00545}{{\ttfamily 1712.00545}}].

\bibitem{Zangeneh2017}
M.~Kord~Zangeneh, A.~Dehyadegari, A.~Sheykhi and R.~B. Mann, \emph{{Microscopic
  Origin of Black Hole Reentrant Phase Transitions}},
  \href{https://doi.org/10.1103/PhysRevD.97.084054}{\emph{Phys. Rev.}
  {\bfseries D97} (2018) 084054}
  [\href{https://arxiv.org/abs/1709.04432}{{\ttfamily 1709.04432}}].

\bibitem{Wei:2019ctz}
S.-W. Wei and Y.-X. Liu, \emph{{Intriguing microstructures of five-dimensional
  neutral Gauss-Bonnet AdS black hole}},
  \href{https://doi.org/10.1016/j.physletb.2020.135287}{\emph{Phys. Lett.}
  {\bfseries B803} (2020) 135287}
  [\href{https://arxiv.org/abs/1910.04528}{{\ttfamily 1910.04528}}].

\bibitem{Kumara:2019xgt}
A.~N. Kumara, C.~L.~A. Rizwan, D.~Vaid and K.~M. Ajith, \emph{{Critical
  Behaviour and Microscopic Structure of Charged AdS Black Hole with a Global
  Monopole in Extended and Alternate Phase Spaces}},
  \href{https://arxiv.org/abs/1906.11550}{{\ttfamily 1906.11550}}.

\bibitem{Kumara:2020mvo}
A.~N. Kumara, C.~L.~A. Rizwan, K.~Hegde, A.~K. M. and M.~S. Ali,
  \emph{{Microstructure and continuous phase transition of a regular Hayward
  black hole in anti-de Sitter spacetime}},
  \href{https://arxiv.org/abs/2003.00889}{{\ttfamily 2003.00889}}.

\bibitem{Xu:2019nnp}
Z.-M. Xu, B.~Wu and W.-L. Yang, \emph{{The fine micro-thermal structures for
  the Reissner-Nordstr\"{o}m black hole}},
  \href{https://arxiv.org/abs/1910.03378}{{\ttfamily 1910.03378}}.

\bibitem{Chabab2018}
M.~Chabab, H.~El~Moumni, S.~Iraoui, K.~Masmar and S.~Zhizeh, \emph{{More
  Insight into Microscopic Properties of RN-AdS Black Hole Surrounded by
  Quintessence via an Alternative Extended Phase Space}},
  \href{https://doi.org/10.1142/S0219887818501712}{\emph{Int. J. Geom. Meth.
  Mod. Phys.} {\bfseries 15} (2018) 1850171}
  [\href{https://arxiv.org/abs/1704.07720}{{\ttfamily 1704.07720}}].

\bibitem{Deng2017}
G.-M. Deng and Y.-C. Huang, \emph{{$Q$-$\Phi$ criticality and microstructure of
  charged AdS black holes in $f(R)$ gravity}},
  \href{https://doi.org/10.1142/S0217751X17502049}{\emph{Int. J. Mod. Phys.}
  {\bfseries A32} (2017) 1750204}
  [\href{https://arxiv.org/abs/1705.04923}{{\ttfamily 1705.04923}}].

\bibitem{Miao2019a}
Y.-G. Miao and Z.-M. Xu, \emph{{Microscopic structures and thermal stability of
  black holes conformally coupled to scalar fields in five dimensions}},
  \href{https://doi.org/10.1016/j.nuclphysb.2019.03.015}{\emph{Nucl. Phys.}
  {\bfseries B942} (2019) 205}
  [\href{https://arxiv.org/abs/1711.01757}{{\ttfamily 1711.01757}}].

\bibitem{Chen2019}
Y.~Chen, H.~Li and S.-J. Zhang, \emph{{Microscopic explanation for black hole
  phase transitions via Ruppeiner geometry: Two competing factors–the
  temperature and repulsive interaction among BH molecules}},
  \href{https://doi.org/10.1016/j.nuclphysb.2019.114752}{\emph{Nucl. Phys.}
  {\bfseries B948} (2019) 114752}
  [\href{https://arxiv.org/abs/1812.11765}{{\ttfamily 1812.11765}}].

\bibitem{Du2019}
Y.-Z. Du, R.~Zhao and L.-C. Zhang, \emph{{Microstructure and Continuous Phase
  Transition of the Gauss-Bonnet AdS Black Hole}},
  \href{https://arxiv.org/abs/1901.07932}{{\ttfamily 1901.07932}}.

\bibitem{Dehyadegari2017}
A.~Dehyadegari, A.~Sheykhi and A.~Montakhab, \emph{{Critical behavior and
  microscopic structure of charged AdS black holes via an alternative phase
  space}}, \href{https://doi.org/10.1016/j.physletb.2017.02.064}{\emph{Phys.
  Lett.} {\bfseries B768} (2017) 235}
  [\href{https://arxiv.org/abs/1607.05333}{{\ttfamily 1607.05333}}].

\bibitem{Ghosh:2019pwy}
A.~Ghosh and C.~Bhamidipati, \emph{{Thermodynamic geometry for charged
  Gauss-Bonnet black holes in AdS spacetimes}},
  \href{https://doi.org/10.1103/PhysRevD.101.046005}{\emph{Phys. Rev.}
  {\bfseries D101} (2020) 046005}
  [\href{https://arxiv.org/abs/1911.06280}{{\ttfamily 1911.06280}}].

\bibitem{Ghosh:2020kba}
A.~Ghosh and C.~Bhamidipati, \emph{{Thermodynamic geometry and interacting
  microstructures of BTZ black holes}},
  \href{https://arxiv.org/abs/2001.10510}{{\ttfamily 2001.10510}}.

\bibitem{Ruppeinerb2008}
G.~Ruppeiner, \emph{{Thermodynamic curvature and phase transitions in
  Kerr-Newman black holes}},
  \href{https://doi.org/10.1103/PhysRevD.78.024016}{\emph{Phys. Rev.}
  {\bfseries D78} (2008) 024016}
  [\href{https://arxiv.org/abs/0802.1326}{{\ttfamily 0802.1326}}].

\bibitem{Ruppeiner95}
G.~Ruppeiner, \emph{{Riemannian geometry in thermodynamic fluctuation theory}},
  \href{https://doi.org/10.1103/RevModPhys.67.605}{\emph{Rev. Mod. Phys.}
  {\bfseries 67} (1995) 605}.

\bibitem{Janyszek_1990}
H.~Janyszek and R.~Mrugaa, \emph{Riemannian geometry and stability of ideal
  quantum gases},
  \href{https://doi.org/10.1088/0305-4470/23/4/016}{\emph{Journal of Physics A:
  Mathematical and General} {\bfseries 23} (1990) 467}.

\bibitem{Oshima_1999x}
H.~Oshima, T.~Obata and H.~Hara, \emph{Riemann scalar curvature of ideal
  quantum gases obeying gentile's statistics},
  \href{https://doi.org/10.1088/0305-4470/32/36/302}{\emph{Journal of Physics
  A: Mathematical and General} {\bfseries 32} (1999) 6373}.

\bibitem{Mirza2008}
B.~Mirza and H.~Mohammadzadeh, \emph{Ruppeiner geometry of anyon gas},
  \href{https://doi.org/10.1103/PhysRevE.78.021127}{\emph{Phys. Rev. E}
  {\bfseries 78} (2008) 021127}.

\bibitem{PhysRevE.88.032123}
H.-O. May, P.~Mausbach and G.~Ruppeiner, \emph{Thermodynamic curvature for
  attractive and repulsive intermolecular forces},
  \href{https://doi.org/10.1103/PhysRevE.88.032123}{\emph{Phys. Rev. E}
  {\bfseries 88} (2013) 032123}.

\bibitem{Bardeen1973}
J.~M. Bardeen, B.~Carter and S.~W. Hawking, \emph{{The Four laws of black hole
  mechanics}}, \href{https://doi.org/10.1007/BF01645742}{\emph{Commun. Math.
  Phys.} {\bfseries 31} (1973) 161}.

\bibitem{AyonBeato:1998ub}
E.~Ayon-Beato and A.~Garcia, \emph{{Regular black hole in general relativity
  coupled to nonlinear electrodynamics}},
  \href{https://doi.org/10.1103/PhysRevLett.80.5056}{\emph{Phys. Rev. Lett.}
  {\bfseries 80} (1998) 5056}
  [\href{https://arxiv.org/abs/gr-qc/9911046}{{\ttfamily gr-qc/9911046}}].

\bibitem{AyonBeato:2000zs}
E.~Ayon-Beato and A.~Garcia, \emph{{The Bardeen model as a nonlinear magnetic
  monopole}}, \href{https://doi.org/10.1016/S0370-2693(00)01125-4}{\emph{Phys.
  Lett.} {\bfseries B493} (2000) 149}
  [\href{https://arxiv.org/abs/gr-qc/0009077}{{\ttfamily gr-qc/0009077}}].

\bibitem{Hayward:2005gi}
S.~A. Hayward, \emph{{Formation and evaporation of regular black holes}},
  \href{https://doi.org/10.1103/PhysRevLett.96.031103}{\emph{Phys. Rev. Lett.}
  {\bfseries 96} (2006) 031103}
  [\href{https://arxiv.org/abs/gr-qc/0506126}{{\ttfamily gr-qc/0506126}}].

\bibitem{A.:2019mqv}
N.~K. A., C.~L.~A. Rizwan, A.~K. M. and M.~S. Ali, \emph{{Photon Orbits and
  Thermodynamic Phase Transition of Regular AdS Black Holes}},
  \href{https://arxiv.org/abs/1912.11909}{{\ttfamily 1912.11909}}.

\bibitem{Fan:2016hvf}
Z.-Y. Fan and X.~Wang, \emph{{Construction of Regular Black Holes in General
  Relativity}}, \href{https://doi.org/10.1103/PhysRevD.94.124027}{\emph{Phys.
  Rev.} {\bfseries D94} (2016) 124027}
  [\href{https://arxiv.org/abs/1610.02636}{{\ttfamily 1610.02636}}].

\bibitem{Fan:2016rih}
Z.-Y. Fan, \emph{{Critical phenomena of regular black holes in anti-de Sitter
  space-time}},
  \href{https://doi.org/10.1140/epjc/s10052-017-4830-9}{\emph{Eur. Phys. J.}
  {\bfseries C77} (2017) 266}
  [\href{https://arxiv.org/abs/1609.04489}{{\ttfamily 1609.04489}}].

\end{thebibliography}\endgroup

\end{document}